\documentclass[12pt]{article}
\usepackage{euscript,amsmath, amssymb, amsfonts}
\usepackage{color}

\newcommand{ \supp}{{\rm supp\,}}
\newcommand{\R}{{\mathbb R}}
\newcommand{ \nn}{\nonumber \\}
\newcommand{ \bee}{\begin{eqnarray}}
\newcommand{ \eee}{\end{eqnarray}}

\newcommand{\p}{{\hbar}}

\newcommand{\oC}{{\mathbb C}}

\newcommand{\di}{{ d }}
\newcommand{\D}{{\cal D}}
\newcommand{\eE}{{\cal E}}

\newcommand{\G}{{\mathbb G}}
\newcommand{\Z}{{\mathbb Z}}
\newcommand{\K}{{\mathbb K}}
\newcommand{\be}{\begin{equation}}
\newcommand{\ee}{\end{equation}}

\newcounter{theorem}
\newcommand{\theorem}{\par\refstepcounter{theorem}
           {\bf Theorem \arabic{section}.\arabic{theorem}. }}
\renewcommand\thetheorem{\thesection.\arabic{theorem}}
\makeatletter \@addtoreset{theorem}{section}

\newcounter{lemma}
\newcommand{\lemma}{\par\refstepcounter{lemma}
           {\bf Lemma \arabic{section}.\arabic{lemma}. }}
\renewcommand\thelemma{\thesection.\arabic{lemma}}
\makeatletter \@addtoreset{lemma}{section}

\newcounter{proposition}
\newcommand{\proposition}{\par\refstepcounter{proposition}
           {\bf Proposition \arabic{section}.\arabic{proposition}. }}
\renewcommand\theproposition{\thesection.\arabic{proposition}}
\makeatletter \@addtoreset{proposition}{section}

\makeatletter \@addtoreset{equation}{section}
\def\theequation{\thesection.\arabic{equation}}
\makeatother

\newcounter{appen}
\newcommand{\appen}[1]{\par\refstepcounter{appen}
{\par\medskip\noindent\Large\bf Appendix \arabic{appen}.
\medskip
}{\Large\bf #1}}

\newcounter{subappen}
\newcommand{\subappen}[1]{\par\refstepcounter{subappen}
{\par\vskip 1cm\noindent\large\bf  \arabic{appen}.\arabic{subappen}.
}{\large\bf #1}\vskip 0.5 cm}
\makeatletter \@addtoreset{subappen}{appen}

\newcounter{subsubappen}

\makeatletter \@addtoreset{subsubappen}{subappen}

\font\frtnfr=eufm10   scaled\magstep1
\font\twlfr=eufm10
\font\tenfr=eufm10

\newfam\frfam
\textfont\frfam=\frtnfr
\scriptfont\frfam=\twlfr
\scriptscriptfont\frfam=\tenfr

\font\frtnopen=msbm10  scaled\magstep2
\font\twlopen=msbm10
\font\tenopen=msbm10

\newfam\openfam
\textfont\openfam=\frtnopen
\scriptfont\openfam=\twlopen
\scriptscriptfont\openfam=\tenopen

\font\frtnsf = cmss12 scaled\magstep1
\font\twlsf = cmss10
\font\tensf = cmss9

\newfam\Scfam
\textfont\Scfam = \frtnsf
\scriptfont\Scfam = \twlsf
\scriptscriptfont\Scfam = \tensf

\begin{document}

%\mark{ DAntiPoisson \ddatt}

\sloppy \title
 {
The deformations of antibracket
with even and odd deformation parameters}
\author
 {
 S.E.Konstein\thanks{E-mail: konstein@lpi.ru}\ \ and
 I.V.Tyutin\thanks{E-mail: tyutin@lpi.ru}
 \thanks{
               This work was supported
               by the RFBR (grant No.~08-02-01118)
 } \\
               {\sf \small I.E.Tamm Department of
               Theoretical Physics,} \\ {\sf \small P. N. Lebedev Physical
               Institute,} \\ {\sf \small 119991, Leninsky Prospect 53,
               Moscow, Russia.} }
\date {
%\ddatt
}

\maketitle

\maketitle

\begin{abstract}
{ \footnotesize We consider antiPoisson superalgebras realized on the smooth
Grassmann-valued functions with compact supports in $\R^n$ and with the grading inverse
to Grassmanian parity. The deformations with even and odd deformation parameters
 of these superalgebras
are presented for arbitrary $n$.
}
\end{abstract}

%%%%%%%%%%%%%%%%%%%%%%%%%%%%%%%%%%%%%%%%%%%%%%%%%%%
\section{Introduction}

In \cite{poi} we described the deformation of Poisson superalgebra depending on even and finite number of
odd deformation parameters. The number of finite parameters in that case may be arbitrary because Poisson superalgebra
realized on the smooth Grassmann-valued functions with compact support has
infinite number of odd 2-cocycles in adjoint representation.

Here we consider the deformations of antiPoisson superalgebras realized on the smooth
Grassmann-valued functions with compact supports in $\R^n$ and show that there exists either
one deformation with one even deformation parameter, or one deformation with one odd parameter.

All necessary definitions are in the next section.
This text organized as follows.

Section \ref{cohom-res} contains previously known results about second cohomology space
of antibracket and two more cohomologies for $n=1$. Theorem \ref{thmain} described the general form
of the deformations are formulated in Section \ref{def-res}
and proved in Section \ref{evoddef}. Cohomology space $H^2_{\mathbf E}$ is described in Section \ref{coh1}
and with details in Appendix \ref{appproof}.

\section{General}

The odd Poisson bracket play an important role in Lagrangian formulation of
the quantum theory of the gauge fields, which is known as BV-formalism
\cite{BV1}, \cite{BV2} (see also \cite{reports}-\cite{HenTei}). These odd
bracket were introduced in physical literature in \cite{BV1} and were called
there as "antibracket". Antibracket possesses many features analogous to ones
of even Poisson bracket and even can be obtained via "canonical formalism" %
with odd time. However, contrary to the case of even Poisson bracket where
there exists voluminous literature on different aspects of the deformation
(quantization) of Poisson algebra, the problem of the deformation of
antibracket is not study satisfactory yet.

In \cite{Leites} the deformations
antibracket  realized on the space of vector fields with
polynomial coefficients are found and in \cite{JMP}
the deformation of antibracket realized on the smooth Grassmann-valued functions with
compact support is found.

The goal of present work is finding all the deformations depending on even and odd
deformation parameters
of antiPoisson superalgebra realized on
the smooth Grassmann-valued functions with
compact supports in $\R^n$.

Let $\K$ be either $\R$ or $\oC$. We denote by ${\cal D}(\R^n)$ the space
of smooth $\K$-valued functions with compact supports on $\R^n$. This space
is endowed with its standard topology. We set
$$
 \mathbf D_{n}= {\cal
D}(\R^{n})\otimes \G^{n},\quad \mathbf E_{n}=
C^\infty(\R^{n})\otimes \G^{},\quad \mathbf D^{\prime }_{n}=
{\cal D}'(\R^{n})\otimes \G^{n},
$$
 where $\G^{n}$ is the Grassmann
algebra with $n$ generators and
$\EuScript D'(\R^{n})$ is the space of
continuous linear functionals on $\EuScript D(\R^{n})$. The generators of
the Grassmann algebra (resp., the coordinates of the space $\R^{n}$) are
denoted by $\xi^\alpha$, $\alpha=1,\ldots,n$ (resp., $x^i$, $i=1,\ldots,
n$).
We shall also use collective variables $z^A$ which are equal to $x^A$
for $A=1,\ldots,n$ and are equal to $\xi^{A-n}$ for
$A=n+1,\ldots,2n$.

The spaces $\mathbf D_{n}$, $\mathbf
E_{n}$, and $\mathbf D^{\prime }_{n}$ possess a natural grading
which is determined by that of the Grassmann algebra.
The Grassmann parity ($\varepsilon$-parity) of an
element $f$ of these spaces is denoted by $\varepsilon(f)$.

The spaces $\mathbf D_{n}$, $\mathbf
E_{n}$, and $\mathbf D^{\prime }_{n}$ possess also another $\Z_2$-grading
$\epsilon$ ($\epsilon$-parity),
 which is inverse to $\varepsilon$-parity: $\epsilon=\varepsilon+1$.

We set
$\varepsilon_A=0$, $\epsilon_A=1$ for $A=1,\ldots, n$
and $\varepsilon_A=1$, $\epsilon_A=0$ for
$A=n+1,\ldots, 2n$.

It is well known, that the bracket
\bee\label{Sch}
[f,g](z)=\sum_{i=1}^n\left(f(z)\frac{\overleftarrow{\partial}}{\partial x^i}
\frac\partial{\partial\xi^i}g(z)-
f(z)\frac{\overleftarrow{\partial}}{\partial\xi^i}
\frac{\partial}{\partial x^i}g(z)\right),
\eee
which we will call ''antibracket'', %
defines the structure of Lie superalgebra on the superspaces
$\mathbf D_{n}$ and
$\mathbf E_{n}$
with the $\epsilon$-parity.

Indeed, $[f,g]=-(-1)^{\epsilon(f)\epsilon(g)}[g,f]$,
$\epsilon([f,g])=\epsilon(f)+\epsilon(g)$,
and Jacobi identity is satisfied:
\be\label{3.0a}
(-1)^{\epsilon(f)\epsilon(h)}[f,[g,h]]
+(-1)^{\epsilon(g)\epsilon(f)}[g,[h,f]]
+(-1)^{\epsilon(h)\epsilon(g)}[h,[f,g]]=0
,\quad f,g,h\in \mathbf
E_{n}.
\ee

Evidently, the
metric $\omega$
defining antibracket
$$
[f,g](z)=f(z)\frac{\overleftarrow{\partial}}{\partial z^A}
\omega^{AB}
\frac\partial{\partial z^B}g(z),
$$
is constant, nondegenerate, and satisfy the condition
$$
\omega^{BA}=-(-1)^{\epsilon_A\epsilon_B}\omega^{AB},\;
\epsilon(\omega^{AB})=\epsilon_A+\epsilon_B,
$$

Here these Lie superalgebras are called antiPoisson
superalgebras.\footnote{
We will consider usual multiplication of the elements of considered antiPoisson
superalgebras with commutation relations $fg=(-1)^{\varepsilon(f)\varepsilon(g)}gf$
as well, and the variables $x^i$ will be called even variables and the variables
$\xi^i$ will be called odd variables.}

The integral on $\mathbf D_{n}$ is defined by the relation $ \int \di z\,
f(z)= \int_{\R^{n}}\di x\int \di\xi\, f(z), $ where the integral on the
Grassmann algebra is normed by the condition $\int \di\xi\,
\xi^1\ldots\xi^{n}=1$. We identify $\G^{n}$ with its dual space $\G^{\prime
n}$ setting $f(g)=\int\di\xi\, f(\xi)g(\xi)$, $f,g\in \G^{n}$.
Correspondingly, the space $\mathbf D^{\prime}_{n}$ of continuous linear
functionals on $\mathbf D_{n}$ is identified  with the space
$\D^\prime(\R^{n})\otimes \G^{n}$. The value $m(f)$ of a functional
$m\in\mathbf D^\prime_n$ on a test function $f\in\mathbf D_n$ will be often
written in the integral form: $m(f)=\int\di z\,m(z)f(z)$.

%%%%%%%%%%%%-----------------------------------------------

%%%%%%%%%%%%-----------------------------------------------

%%%%%%%%%%%%-----------------------------------------------

\section{Cohomology of antibrackets (Results)}\label{cohom-res}

Let $\mathbf D_{n}$ acts in a $\Z_2$-graded space $V$ (the action of $f\in
\mathbf D_{n}$ on $v\in V$ will be denoted by $f\cdot v$). The space
$C_p(\mathbf D_{n},V)$ of $p$-cochains consists of all multilinear
superantisymmetric mappings from $\mathbf D_{n}^p$ to $V$. Superantisymmetry
means, as usual, that $M_p(\,...\,,f_i,\,f_{i+1},...)=-
(-1)^{\epsilon(f_i)\epsilon(f_{i+1})}M_p(\,...,f_{i+1},f_{i},...)$. The space
$C_p(\mathbf D_{n}, V)$ possesses a natural $\Z_2$-grading: by definition,
$M_p\in C_p(\mathbf D_{n},V)$ has the definite $\epsilon$-parity
$\epsilon_{M_p}$ if
$$
\epsilon(M_p(f_1,\ldots,f_p))=
\epsilon_{M_p}+\epsilon(f_1)+\ldots+\epsilon(f_p)
$$
for any $f_j\in\mathbf D_{n}$ with $\epsilon$-parities $\epsilon(f_j)$. We
will often use the Grassmann $\varepsilon$-parity\footnote{If $V$ is the
space of Grassmann-valued functions on $\R^n$ then $\varepsilon$ defined in
such a way coincides with usual Grassmann parity. } of cochains:
$\varepsilon_{M_p}=\epsilon_{M_p}+p+1$.
The differential $\di_p^V$ is defined to be
the linear operator from $C_p(\mathbf D_{n}, V)$ to $C_{p+1}(\mathbf D_{n}, V)$
such that
\begin{eqnarray}
&&d_p^{V}M_p(f_1,...,f_{p+1})=
-\sum_{j=1}^{p+1}(-1)^{j+\epsilon(f_j)|\epsilon(f)|_{1,j-1}+
\epsilon(f_j)\epsilon_{M_p}}f_j\cdot
M_p(f_{1},...,\breve{f}_j,...,f_{p+1})- \nonumber \\
&&-\sum_{i<j}(-1)^{j+\epsilon(f_j)|\epsilon(f)|_{i+1,j-1}}
M_p(f_1,...f_{i-1},[f_i,f_j],f_{i+1},...,
\breve{f}_j,...,f_{p+1}),\label{diff}
\end{eqnarray}
for any $M_p\in C_p(\mathbf D_{n}, V)$ and $f_1,\ldots,f_{p+1}\in\mathbf
D_{n}$ having definite $\epsilon$-parities. Here the sign $\breve{}$ means
that the argument is omitted and the notation
$$
|\epsilon(f)|_{i,j}=\sum_{l=i}^j\epsilon(f_l)
$$
 has been used.
 The
differential $\di^V$ is nilpotent (see \cite{Schei97}), {\it i.e.},
$\di^V_{p+1}\di^V_p=0$ for any $p=0,1,\ldots$. The $p$-th cohomology space of
the differential $\di_p^V$ will be denoted by $H^p_V$. The second cohomology
space $H^2_{\mathrm{ad}}$ in the adjoint representation is closely related to
the problem of finding formal deformations of the Lie bracket $[\cdot,\cdot]$
of the form $ [f,g]_*=[f,g]+\hbar[f,g]_1+\ldots$ up to
similarity transformations $[f,g]_T=T^{-1}[Tf,Tg]$ where continuous
linear operator $T$
from $V[[\p]]$ to $V[[\p]]$ has the form
$T=\mathsf {id}+\p T_1$.

The condition that
$[\cdot,\cdot]_1$ is a 2-cocycle is equivalent to the Jacobi identity for
$[\cdot,\cdot]_*$ modulo the $\hbar$-order terms.

In the present paper, similarly to \cite{SKT1}, we suppose that cochains are
separately continuous multilinear mappings.

We need the cohomologies of the antiPoisson algebra
$\mathbf D_{n}$ in the
following representations:
\begin{enumerate}
\item
$V=\mathbf E_{n}$ and $f\cdot g=[f,g]$ for any $f\in\mathbf D_{n}$,
$g\in\mathbf E_{n}$. The space $C_p(\mathbf D_{n}, \mathbf E_{n})$ consists
of separately continuous superantisymmetric multilinear mappings from $(\mathbf
D_{n})^p$ to $\mathbf E_{n}$. The cohomology spaces and the differentials
will be denoted by $H^p_{\mathrm{E}}$ and $\di^{\mathrm{ad}}_p$ respectively.
\item
The adjoint representation: $V=\mathbf D_{n}$ and $f\cdot g=[f,g]$ for any
$f,g\in\mathbf D_{n}$. The space $C_p(\mathbf D_{n}, \mathbf D_{n})$ consists
of separately continuous superantisymmetric multilinear mappings from $(\mathbf
D_{n})^p$ to $\mathbf D_{n}$. The cohomology spaces and the differentials
will be denoted by $H^p_{\mathrm{ad}}$ and $\di^{\mathrm{ad}}_p$
respectively.
\end{enumerate}
We shall call
p-cocycles $M_p^1,\ldots M_p^k$ independent cohomologies if they give rise to linearly
independent elements in $H^p$. For a multilinear form $M_p$ taking values in
$\mathbf D_{n}$, $\mathbf E_{n}$, or $\mathbf D^{\prime}_{n}$, we write
$M_p(z|f_1,\ldots,f_p)$ instead of more cumbersome $M_p(f_1,\ldots,f_p)(z)$.

The following theorems proved in \cite{JMP} describe these cohomology of antibracket

\theorem\label{th2}{}
{\it
Let the bilinear mappings $m_{2|1}$, $m_{2|2}$, $m_{2|5}$,
$m_{2|6}$
from $(\mathbf D_1)^2$ to
$\mathbf E_{1}$ and bilinear mappings $m_{2|3}$, $m_{2|4}$ from
$(\mathbf D_n)^2$ to $\mathbf E_{n}$ be defined by the relations
\begin{eqnarray}
&&m_{2|1}(z|f,g)=\int du \partial_\eta g(u)\partial^{3}_y f(u),\;\;
\epsilon_{m_{2|1}}=1, \label{5.2.7} \\
&&m_{2|2}(z|f,g)=\int du\theta(x-y)[\partial_\eta g(u)\partial_y^3f(u)-
\partial_\eta f(u)\partial_y^3 g(u)]+ \nonumber \\
&&+x[\{\partial_\xi\partial_{x}^{2}f(z)\}\partial_\xi\partial_{x}g(z)-
\{\partial_\xi\partial_xf(z)\}\partial_\xi
\partial_x^2g(z)],\;\; \epsilon_{m_{2|2}}=1,
\label{5.2.8} \\
&&m_{2|3}(z|f,g)=(-1)^{\varepsilon(f)}\{(1-N_{\xi})f(z)\}(1-N_{\xi})g(z), \;\;
\epsilon_{m_{2|3}}=1, \label{6.3a} \\
&&m_{2|4}(z|f,g)=(-1)^{\varepsilon (f)}\{\Delta f(z)\}\eE_z g(z)+
\{\eE_z f(z)\}\Delta g(z)\;\;  \epsilon_{m_{2|4}}=0. \label{6.3b}\\
&&m_{2|5}(z|f,g)=\int du (-1)^{\epsilon(f)}\partial_y f(u)\partial_y g(u),\;\;
\epsilon_{m_{2|5}}=0, \label{5.2.7-5} \\
&&m_{2|6}(z|f,g)=\int du\theta(x-y)(-1)^{\epsilon(f)}\partial_y f(u)\partial_y g(u),\;\;
\epsilon_{m_{2|6}}=0 \label{5.2.7-6}
\end{eqnarray}
where $z=(x,\xi)$, $u=(y,\eta)$, $N_\xi=\xi \partial_{\xi}$, and
\begin{equation}
\Delta =\partial_x \partial_\xi. \label{Delta}
\end{equation}

Then

\begin{enumerate}

\item\label{itemad}
$H^2_{\mathrm {ad}}\simeq \K^2$ and the cochains
$m_{2|3}(z|f,g)$ and $m_{2|4}(z|f,g)$
are
independent nontrivial cocycles.

\item\label{item}
Let $n=1$.

Then
$H^2_{\mathbf E}\simeq  \K^6$ and the cochains
$m_{2|1}(z|f,g)$, $m_{2|2}(z|f,g)$, $m_{2|3}(z|f,g)$, $m_{2|4}(z|f,g)$, $m_{2|5}(z|f,g)$, and $m_{2|6}(z|f,g)$
are
independent nontrivial cocycles.

\item
Let $n\geq 2$. Then $H^2_{\mathbf D'}\simeq H^2_{\mathbf E}\simeq \K^2$ and
the cochains $m_{2|3}(z|f,g)$ and $m_{2|4}(z|f,g)$ are independent nontrivial
cocycles.

\end{enumerate}
}

Nonlocal cocycles $m_{2|5}$ and $m_{2|6}$ are lost in \cite{JMP} for $n=1$. This is the reason to reproduce
below (in Section \ref{coh1}
and in Appendix \ref{appproof}) the proof of item \ref{item} and item \ref{itemad} for $n=1$ of Theorem \ref{th2}.

%%%%%%%%%%%%-----------------------------------------------

%%%%%%%%%%%%-----------------------------------------------

%%%%%%%%%%%%-----------------------------------------------

\section{Deformations of antibrackets (Results)}\label{def-res}

Consider general form of deformation, $[f,g]_{\ast }(z)$, of the antibracket
$[f,g](z)$.

Because antibracket on $\mathbf D_n$ has only two independent adjoint second cohomology, one even ($m_{2|4}$)
and one odd ($m_{2|3}$), we consider the deformations depending on one even and one odd (in Grassmannian sense)
parameters, $\hbar$
and $\theta$ correspondingly.

We will suppose that:

\begin{enumerate}

\item
\begin{eqnarray}
&& [f,g]_*(z)\equiv C(z|f,g;\hbar, \theta)=A_1(z|f,g;\hbar) + \theta A_{0}(z|f,g;\hbar), \ \ \epsilon(A_i)=i+1\\
 \label{decc}\\
&& A_{1}(z|f,g;0)=\lbrack f,g\rbrack (z), \notag
\end{eqnarray}

\item
\begin{equation*}
A_{i}(z|f,g;\hbar)=\sum_k \hbar^k A_{i|k}(z|f,g)
\end{equation*}

\item
$A_{i|k}(z|f,g)\in \mathbf D_n$,
for all $f,g \in \mathbf D_n$;

\item
 $[f,g]_{\ast }(z)$ satisfies the Jacobi identity
\begin{equation}
(-1)^{\epsilon (f)\epsilon (h)}[[f,g]_{\ast },h]_{\ast }+
\mathrm{cycle}(f,g,h)=0,\;\forall f,g,h \in \mathbf D_n,
\label{1.1}
\end{equation}
or
\begin{equation}
(-1)^{\epsilon (f)\epsilon (h)}C(z|C(|f,g),h)+\mathrm{cycle}
(f,g,h)=0.  \label{1.1a}
\end{equation}

\end{enumerate}

Note that if a form $C(z|f,g)$ satisfies the Jacobi identity then the form
$C_{T}(z|f,g)$,
\begin{equation*}
C_{T}(z|f,g)=T^{-1}C(z|Tf,Tg),
\end{equation*}
satisfies the Jacobi identity too. Here $T$: $f(z)\rightarrow T(z|f)$ is
invertible continuous map $\mathbf D_n\rightarrow \mathbf D_n$.

Formal deformations $C^1$ and $C^2$ are called similar if there is a
continuous $\K[[\hbar, \theta]]$-linear parity conserving similarity operator
$T: \mathbf D_{n}[[\hbar, \theta]]\to\mathbf D_{n}[[\hbar, \theta]]$ such that
$TC^1(f,g)=C^2(T f,Tg)$, $f,g\in \mathbf D_n[[\hbar, \theta]]$ and $T=id + T_1$,
where $T_1=0$ if $\hbar=0$ and $\theta=0$.

Theorem \ref{th2} allows us to prove the following theorem, stating the
general form of the deformation of antiPoisson superalgebra with even deformation parameter:

\theorem\label{theven}{}\cite{JMP}
{\it The deformation of antiPoisson superalgebra with even parameter $\hbar$
has the form
\begin{equation}\label{even}
[f(z),\, g(z)]_*\!=\![f(z),g(z)] + (-1)^{\varepsilon(f)}\{\frac{\hbar
c}{1+\hbar cN_z/2}\Delta f(z)\} \eE_zg(z)
+
\{\eE_z f(z)\}
\frac{\hbar c}{1+\hbar cN_z/2}\Delta g(z)
\end{equation}
up to similarity transformation,
where $N_z=z^A \frac {\partial}{\partial z^A}$, and $c$ is an arbitrary
formal series
in $\hbar$ with coefficients in $\K$.}

The identity $\theta^2=0$ and Theorem \ref{th2} lead to evident result:

\theorem\label{thodd}{}
{\it The deformation of antiPoisson superalgebra with odd parameter $\theta$
has the form
\begin{equation}\label{odd}
[f(z),\, g(z)]_*\!=\![f(z),g(z)] + \theta((-1)^{\varepsilon (f)}\{\Delta f(z)\}\eE_z g(z)+
\{\eE_z f(z)\}\Delta g(z))
\end{equation}
}

Main result of present work is the following theorem which is proved below

\theorem\label{thmain}{}
{\it The deformation of antiPoisson superalgebra with one even and odd parameters
has either the form (\ref{even}) or the form (\ref{odd})}.

%%%%%%%%%%%%-----------------------------------------------

%%%%%%%%%%%%-----------------------------------------------

%%%%%%%%%%%%-----------------------------------------------

\section{Preliminary and Notation}\label{not}

We define $\delta$-function by the formula
\[
\int dz^{\prime }\delta (z^{\prime }-z)f(z^{\prime })=\int f(z^\prime)
\delta (z-z^{\prime })dz^{\prime }=f(z).
\]

Evidently,
\[
[f,g](z)=(-1)^{\varepsilon_A\epsilon(f)}\frac{\partial}{\partial z^A}
(f(z)\omega^{AB}\frac{\partial}{\partial z^B}g(z))-2f\Delta g(z),\]
\[
(-1)^{\epsilon(g)}\int dzf[g,h]=\int dz[f,g]h+2\int dzf\Delta gh,
\]
where $\Delta$ is defined by (\ref{Delta}).

The following notation is used below:
\begin{eqnarray}
&& T_{\ldots (A)_{k}\ldots }\equiv T_{\ldots A_{1}\ldots A_{k}\ldots },\quad
T_{\ldots A_{i}A_{i+1}\ldots }=(-1)^{\varepsilon _{A_{i}}\varepsilon
_{A_{i+1}}}T_{\ldots A_{i+1}A_{i}\ldots },\quad i=1,\ldots ,k-1 \nn
&& T_{\ldots (A)_{k}\ldots }Q_{\ldots }{}^{(A)_{k}}{}_{\ldots }\equiv T_{\ldots
A_{1}\ldots A_{k}\ldots }Q_{\ldots }{}^{A_{1}\ldots A_{k}}{}_{\ldots }, \nn
&& (\partial_A)^Q\equiv \partial_{A_1}\partial_{A_2}\ldots
\partial_{A_Q},\ \  (p_A)^Q\equiv p_{A_1}p_{A_2}\ldots p_{A_Q},
\notag
\end{eqnarray}
and so on.

We denote by $M_p(\ldots)$ the separately continuous
superantisymmetrical $p$-linear forms on
$(\mathbf D_{n})^p$. Thus, the arguments of these functionals are the
functions $f(z)$ of the form
\begin{equation}\label{dec}
f(z)=\sum_{k=0}^{n} f_{(\alpha)_k}(x)(\xi^\alpha)^k \in {\mathbf D}_n,
\quad f_{(\alpha)_k}(x)\in \D(\R^{n}).
\end{equation}
For any $f(z)\in {\mathbf D}_n$ we can define the support
$$ \mathrm{supp}(f)\stackrel {def} = \bigcup_{(\alpha)_k}
\mathrm{supp}(f_{(\alpha)_k}(x)).$$
For each set $V\subset \R^n$ we use the notation $z\bigcap V=\varnothing$
if $z=(x,\xi)$ and there exist some domain $U\subset \R^n$ such that
 $x\in U$ and $U\bigcap V=\varnothing$.

It can be easily proved
that such multilinear forms can be written in the integral form (see
\cite{SKT1}):
\begin{equation} M_p(f_1,\ldots,f_p)=\int dz_p\cdots
dz_1m_p(z_1,\ldots,z_p) f_1(z_1)\cdots f_p(z_p),\;p=1,2,... \label{3.1n}
\end{equation}
and
\begin{equation} M_p(z|f_1,\ldots,f_p)=\int dz_p\cdots
dz_1m_p(z|z_1,\ldots,z_p) f_1(z_1)\cdots
f_p(z_p),\;p=1,2,\,...\,\,.\label{3.2n}
\end{equation}
Let by definition
\[
\epsilon(M_p(f_1,\ldots,f_p))
=\epsilon_{m_p}+pn+
\epsilon(f_1)+\ldots+\epsilon(f_p).
\]
It follows from the properties of the forms $M_p$ that the corresponding
kernels $m_p$ have the following properties:
\begin{eqnarray}
&&\epsilon_{m_p}=pn+\epsilon_{M_p},\;\;\varepsilon_{m_p}=pn+
\varepsilon_{M_p},\;\;\epsilon_{m_p}=\varepsilon_{m_p}+p+1, \nonumber \\
&&m_p(*|z_1\ldots z_i,z_{i+1}\ldots z_p)=(-1)^{n} m_p(*|z_{1}\ldots
z_{i+1}^*,z_i^*\ldots z_p). \label{3.2an}
\end{eqnarray}
Here  $z^*=(x,\,-\xi)$ if $z=(x,\,\xi)$.

Introduce the space ${\cal M}_1\subset {C}_2(\mathbf D_n,\,\mathbf D_n')$
consisting of all 2-forms which can be locally represented as
\begin{eqnarray}
M_{2|2}^{1}(z|f,g) = \sum_{q = 0}^Q
m^{1(A)_q}(z|[(\partial^z_A)^qf(z)]g- (-1)^{\epsilon(f)\epsilon(g)}
[(\partial^z_A)^qg(z)]f),
\label{5.4d}
\end{eqnarray}
with locally constant $Q$ and
the space ${\cal M}_2\subset {C}_2(\mathbf D_n,\,\mathbf D_n')$
consisting of all 2-forms which can be locally represented as
\begin{eqnarray}
M_{2|2}^{2}(z|f,g) = \sum_{q =
0}^Q m^{2(A)_q}(z|[(\partial_A)^qf]g- (-1)^{\epsilon(f)\epsilon(g)}
(\partial_A)^qg]f)
\label{5.4h}
\end{eqnarray}
with locally constant $Q$,
where
$m^{1,2(A)_q}(z|\cdot)\in {C}_1(\mathbf D_n,\,\mathbf D_n')$.

The space ${\cal M}_0 = {\cal M}_1 \bigcap {\cal M}_2$ is called in
this paper the space of local bilinear forms. It consists of all the form,
which can be present as
$$
M_{2|\,\mathrm {loc}}(z|f,g) = \sum_{p,\,q =
0}^Q m^{(A)_q|(B)_p}(z)\left((\partial_A)^q f(z) \, (\partial_B)^p g (z) -
(-1)^{\epsilon(f)\epsilon(g)}
(\partial_A)^q  g(z)\, (\partial_B)^p f (z)
\right).
$$
Here $m^{(A)_q|(B)_p}\in D'\otimes \G^{n} $, and
the summation limit $Q$ is locally constant with respect to $z$.

%%%%%%%%%%%%-----------------------------------------------

%%%%%%%%%%%%-----------------------------------------------

%%%%%%%%%%%%-----------------------------------------------

%%%%%%%%%%%%-----------------------------------------------

%%%%%%%%%%%%-----------------------------------------------

%%%%%%%%%%%%-----------------------------------------------

%%%%%%%%%%%%-----------------------------------------------

%%%%%%%%%%%%-----------------------------------------------

%%%%%%%%%%%%-----------------------------------------------

\section{$H^2_{\mathbf E}$ for $n=1$ antibracket}\label{coh1}

Here we give the proof of the point b) in Theorem \ref{th2}.

\proposition\label{propadn=1}
{\it Let $n=1$.

Let
the bilinear form
\begin{equation*}
M_{2}(z|f,g)=\int dvdum_{2}(z|u,v)f(u)g(v),
\end{equation*}
such that $M_{2}(z|f,g)\in \mathbf E_1$ for all $f,g \in \mathbf D_1$
be cocycle, i.e. it satisfy the cohomology equation
\begin{eqnarray}
&&d_{2}^{\mathrm{ad}}M_{2}(z|f,g,h)=-(-1)^{\epsilon (f)\epsilon (h)}\{(-1)^{\epsilon
(f)\epsilon (h)}[M_{2}(z|f,g),h(z)]+ \nonumber \\
&&+(-1)^{\epsilon (f)\epsilon (h)}M_{2}(z|[f,g],h)+
\mathrm{cycle}(f,g,h)\}=0.  \label{5.1}
\end{eqnarray}

Then
\begin{eqnarray*}
&&M_{2}(z|f,g)=c_{1}m_{2|1}(x|f,g)+c_{2}m_{2|2}(x|f,g)+c_{5}m_{2|5}(x|f,g)+
\\
&&+c_{6}m_{2|6}(x|f,g)+d_{1}^{\mathrm{ad}}M_{1|1}(z|f,g)+M_{2\mathrm{loc}%
}(z|f,g).
\end{eqnarray*}%
where $c_i$ are constants, $m_{2|i}$ are defined in Theorem \ref{th2} and $M_{2\mathrm{loc}%
}(z|f,g) \in {\cal M}_0$.
}

The details of the proof can be found in Appendix \ref{appproof}.

The space of local cocycles is generated up to cobondaries by $m_{2|3}$ (odd cocycle)
and $m_{2|4}$ (even cocycle) \cite{JMP}.

%%%%%%%%%%%%%%%%%%%%%%%%%%%%%%%%%%%
%%%%%%%%%%%%%%%%%%%%%%%%%%%%%%%%%%%

\section{Deformation with one even and one odd parameter}\label{evoddef}

Let%
\begin{eqnarray}
&&\,[f(z),\,g(z)]_{\ast }=A(z|f,g;\hbar ,\theta )=A_{1}(z|f,g;\hbar )+\theta
A_{0}(z|f,g;\hbar ),  \notag \\
&&\,J_{A,A}(z|f,g,h)=(-1)^{(\epsilon (f))(\epsilon
(h))}A(z|A(|f,g;\hbar ,\theta ),h;\hbar ,\theta )+\mathrm{cycle}(f,g,h)=0,
\label{2.1}
\end{eqnarray}%
where $\varepsilon _{\hbar }=0$, $\varepsilon _{A_{1}}=1$, $\varepsilon
_{\theta }=1$, $\varepsilon _{A_{0}}=0$.

It follows from Jacobi
identity (\ref{2.1}):%
\begin{equation}
J_{A_{1},A_{1}}(z|f,g,h)=0,
\end{equation}%
\begin{equation}
J_{A_{1},\theta A_{0}}(z|f,g,h)=0,\label{2odd}
\end{equation}%
such that we have from Theorem \ref{theven}
\begin{equation}
A_1[f, g;\hbar]=[f(z),g(z)] + (-1)^{\varepsilon(f)}\{\frac{\hbar
c}{1+\hbar cN_z/2}\Delta f(z)\} \eE_zg(z)
+
\{\eE_z f(z)\}
\frac{\hbar c}{1+\hbar cN_z/2}\Delta g(z)
\end{equation}
(up to similarity transformation of $[f(z),\,g(z)]_{\ast }$%
)

If $A_0\ne 0$ then we can redefine $\theta\mapsto \hbar^{-k} \theta$
with some definite $k$ in such a way that the decomposition of
$A_0(z|f,g;\hbar)$ starts with zero degree of $\hbar$: $A_0(z|f,g;0)\ne 0$.

Then (\ref{2odd}) gives $J_{A_{1},\theta A_{0}}(z|f,g,0)=0$, i.e.
$A_0|_{\hbar=0}$ is a cocycle, and since it is odd,
$A_0(z|f,g;0)= c_{0|0}m_{2|3}{z|f,g}$ up to equivalence transformation.

To prove Theorem \ref{def-res} it remains to prove that if
$A_0\ne 0$ then $A_1(z|f,g;\hbar)=[f,g]$.

%\subsection{A.}

Let us assume that $A_0\ne 0$.
Then we may assume that
\begin{equation*}
A_0(z|f,g;\hbar)=\sum_{k_0=0}^\infty A_{0|k_0}(z|f,g),\ \ A_{0|0}(z|f,g)\ne 0.
\end{equation*}

Let
\begin{eqnarray*}
&&A_{1}(z|f,g)=[f(z),g(z)]+\hbar
^{k_{1}+1}c_{1|k_{1}}m_{2|4}(z|f,g)+O(\hbar ^{k_{1}+2}).
\end{eqnarray*}

Define the notation
\begin{eqnarray*}
&&A_{0|[m,n]}(z|f,g)=\sum_{l=m}^{n}\hbar ^{l}A_{0|l}(z|f,g).
\end{eqnarray*}

%%%%%%%%%%%%%%%%%%%%%%%%%%%%%%%%%%%%%%%%%%%%%%%%%%%%%%%%%%%%%%%%%%%%%%%%
%%%%%%%%%%%%%%%%%%%%%%%%%%%%%%%%%%%%%%%%
%%%%%%%%%%%%%%%%%%%%%%%%%

%\subsection{B}

Let
\begin{eqnarray*}
&&c_{1}=O(\hbar ^{k_{1}}),
\end{eqnarray*}%
where $k_{1}$ is some integers. $k_{1}\geq 1$.

\subsubsection{$0$-th, ... ,$(k_{1})$-th orders in $\hbar $}

In these cases, we find%
\begin{equation*}
d_{2}^{\mathrm{ad}}A_{0|[0,k_{1}]}(z|f,g,h)=0,
\end{equation*}

such that we obtain (up to similarity transformation)%
\begin{equation*}
A_{0|[0,k_{1}-1]}(z|f,g,h)=c_{0|[0,k_{1}]}m_{2|3}(z|f,g),\;c_{0|0}\neq 0.
\end{equation*}

Here $c_{0|[m,n]}=\sum_{k=m}^n c_{0|k}$, $c_{0|0}\ne 0$.

Before we will start to consider remaining case
let us formulate the following proposition

\proposition\label{prop2}
{\it
Let
\begin{eqnarray}
&&(-1)^{(\epsilon (f))(\epsilon (h))}\left[
[A(z|f,g),h(z)]+A(z|[f,g],h)\right] +%
\mathrm{cycle}(f,g,h)+  \notag \\
&&\,+c J_{m_{2|3},m_{2|4}}(z|f,g,h)=0.  \label{proop2}
\end{eqnarray}
for some $c\in \K$ and some $A\in C_2 (\mathbf D_n,\,\mathbf D_n)$

Then $c=0$.
}

Proof.

1. Note that up to some similarity transformation
$A$ is local form, $A\in {\cal M}_0$.

Indeed,
consider the domains

i) $z\bigcap [\supp(f)\bigcup \supp(g)\bigcup \supp(h)]=
\supp(f)\bigcap [\supp(g)\bigcup \supp(h)]=\varnothing$

and

ii) $z\bigcap [\supp(f)\bigcup \supp(g)\bigcup \supp(h)]=\varnothing$

In these domains,
$J_{m_{2|3},m_{2|4}}(z|f,g,h)=0$ and, as it is shown in \cite{JMP},
$A(z|f,g)$ can be represented in the form

\begin{eqnarray*}
&& A(z|f,g)=A_{\mathrm {loc}}(z|f,g)+d_1^{\mathrm {ad}} M(z|f,g)\\
&& A_{\mathrm {loc}}(z|f,g) = \sum_{a,b=0}^N (-1)^{\varepsilon(f){|\varepsilon_B|_{1,b}+1}}
    m^{(A)_a|(B)_b}(z) [(\partial_A^z)^af(z)](\partial_B^z)^b g(z)\\
&& m^{(B)_b|(A)_a} = (-1)^{|\varepsilon_A|_{1,a} |\varepsilon_B|_{1,b} } m^{(A)_a|(B)_b}
,\ \
\varepsilon(m^{(A)_a|(B)_b} (\partial_A^z)^a(\partial_B^z)^b)=0,
\end{eqnarray*}
where $M(z|f)$ is some 1-form, $\varepsilon_{M}=1$.

After similarity transformation of $[f,g]$ with $T(z|f)=f(z)-\hbar \theta M(z|f)$
we have $A(z|f,g)=A_{\mathrm {loc}}(z|f,g$ and $A_{\mathrm {loc}}(z|f,g$
satisfies eq. (\ref{proop2}).

Choosing $f(z)=e^{zp}$, $g(z)=e^{zq}$, $h(z)=e^{zr}$ in some neighbourhood of $z$, we reduce
eq. (\ref{proop2}) to the form
\begin{eqnarray}
&&\Phi(z|p,q,r)\langle p,q\rangle - [F(z|p,q),zr] + \mathrm{cycle} (p,q,r) =\notag\\
&& = c \cdot \left(
m_{2|3}(z|m_{2|4};p,q,r) + m_{2|4}(z|m_{2|3}; p,q,r) + \mathrm{cycle} (p,q,r)
\right)\label{mainlocal}
\end{eqnarray}
where
\begin{eqnarray*}
F(z|p,q) &=&
\sum_{a,b=0}^N (-1)^{\varepsilon(f){|\varepsilon_B|_{1,b}+1}}
    m^{(A)_a|(B)_b}(z) (p_A)^a (q_B)^b = F(z|q,p)=\\
&& =
m^{0|0}(z) + m^A (z) (p_A + q_A) + O((\mathrm{momenta})^2),\\
&& m^A(z)=m^{0|A}(z)=m^{A|0}(z),\\
\langle p,q\rangle&=&[e^{zp},\, e^{zq}] e^{-z(p+q)}\\
\Phi(z|p,q,r)&=& F(z|p+q,r) - F(z|p,r)-F(z|q,r)=\\
&&=-m^{0|0}(z)+O(\mathrm{momenta})\\
m_{2|3}(z|m_{2|4};p,q,r)
&=&
m_{2|3}(z|m_{2|4}(|e^{zp},e^{zq}),e^{zr})e^{-z(p+q+r)}=\\
&&=
-\frac 1 2 \{ [\langle p,p\rangle(1-zq/2)+\langle q,q\rangle (1-zp/2)](1-\xi\alpha-\xi\beta)+\\
&&+\langle p,p\rangle\xi\beta/2 + \langle q,q\rangle \xi\alpha/2\} (1-\xi\gamma)
\\
m_{2|4}(z|m_{2|3};p,q,r)
&=&
m_{2|4}(z|m_{2|3}(|e^{zp},e^{zq}),e^{zr})e^{-z(p+q+r)}=\\
&&=\left\{
\xi\alpha(\xi\beta-1)(u\alpha+v\alpha)+\xi\beta(\xi\alpha - 1)(u\beta + v\beta)
\right\}(1-zr/2)+\\
&&+\frac 1 2
\left\{
1-\frac 1 2 \xi\alpha - \frac 1 2 \xi\beta  - \frac 1 2 (1-\xi\alpha)(1-\xi\beta)(zp+zq)
\right\}\langle r,r\rangle.
\end{eqnarray*}

Here
$$
zp=z^A p_A,\ p_A=(u_i,\alpha_i),\ q_A=(v_i,\beta_i),\ r_A=(t_i,\gamma_i),
\ \xi\alpha=\xi^i\alpha_i, \ u\alpha=u_i \alpha_i
$$
and so on.

For $r=0$, we find
\begin{eqnarray*}
&&\left[ m_{2|3}(z|m_{2|4};p,q,r)  +  m_{2|4}(z|m_{2|3}; p,q,r) + \mathrm{cycle} (p,q,r)\right] |_{r=0}=P_4(p,q)\\
&&P_4(p,q) =
 -u\alpha-v\beta+(u\alpha)(\xi\beta)+(v\beta)(\xi\alpha)-(u\beta)(\xi\beta)
-(v\alpha)(\xi\alpha)+\\
&& \ \ \ \ \ \ \ \ \ \ \ +(u\beta)(\xi\alpha)(\xi\beta)+(v\alpha)(\xi\alpha)(\xi\beta)
\end{eqnarray*}

At $r=0$, eq. (\ref{mainlocal})
takes the form
\begin{eqnarray}
&&\Phi(z|p,q)\langle p,q \rangle - [F(z|p),\, zq] - [F(z|q),\, zp]=c\, P_4(p,q),
\label{last}
\\
&&
\Phi(z|p,q)=F(z|p+q) - F(z|p) - F(z|q), \ \ F(z|p)=F(z|p,0).\notag
\end{eqnarray}

Consider in eq. (\ref{last}) the terms of the second order in momenta.
We obtain the reduced equation
$$
m^{0|0}(z)\langle p,q \rangle + [m^A (z)q_A,\, zp] +[m^A (z)p_A,\, zq] = c\,(u\alpha+v\beta))
$$
which implies
$$c=0.$$
Q.E.D.

\subsubsection{$(k_{1}+1)$-th order in $\hbar $}

In this case, we find%
\begin{eqnarray}
&&(-1)^{(\epsilon (f))(\epsilon (h))}\left[
[A_{0|k_{1}+1}(z|f,g),h(z)]+A_{0|k_{1}+1}(z|[f,g],h)\right] +%
\mathrm{cycle}(f,g,h)+  \notag \\
&&\,+c_{0|0}c_{1|k_1}J_{m_{2|3},m_{2|4}}(z|f,g,h)=0.  \label{2.2.2.1}
\end{eqnarray}

It follows from eq. (\ref{2.2.2.1}) and Proposition \ref{prop2} that%
\begin{equation*}
c_{0|0}c_{1|k_1}=0.
\end{equation*}
and so $c_{1|k_1}=0$.

Using the induction method, we obtain that if $A_0\ne 0$
then the general solution of eq. (\ref{2.1}) (up to similarity transformation) is%
\begin{equation*}
\,[f(z),\,g(z)]_{\ast
}=\,[f(z),\,g(z)]+\theta A_{0}(z|f,g)=\,[f(z),\,g(z)]+\theta \sum_i c_{0|i} \hbar^i m_{2|3}(z|f,g),
\end{equation*}
or after redefining $\theta$
\begin{equation*}
\,[f(z),\,g(z)]_{\ast}=[f(z),\,g(z)]+\theta m_{2|3}(z|f,g).
\end{equation*}

%%%%%%%%%%%%%%%%%%%%%%%%%%%%%%%%%%%%%%%%%%%%%%%%%%%%%%%%%%%%%%%%%%%%
%%%%%%%%%%%%%%%%%%%%%%%%%%%%%%%%%%%%%%%%%%%%%%%%%%%%%%%%%%%%%%%%%%%%

%%%%%%%%%%%%-----------------------------------------------

%%%%%%%%%%%%-----------------------------------------------

%%%%%%%%%%%%-----------------------------------------------

%%%%%%%%%%%%-----------------------------------------------

\setcounter{MaxMatrixCols}{10}

\def\theequation{A\arabic{appen}.\arabic{equation}}

\renewcommand{\theorem}{\par\refstepcounter{theorem}
{\bf Theorem A\arabic{appen}.\arabic{theorem}. }}
\renewcommand{\lemma}{\par\refstepcounter{lemma}
{\bf Lemma A\arabic{appen}.\arabic{lemma}. }}
\renewcommand{\proposition}{\par\refstepcounter{proposition}
{\bf Proposition A\arabic{appen}.\arabic{proposition}. }}
\makeatletter \@addtoreset{theorem}{appen}
\makeatletter \@addtoreset{lemma}{appen}
\makeatletter \@addtoreset{proposition}{appen}
\makeatletter \@addtoreset{equation}{appen}
\renewcommand\thetheorem{A\theappen.\arabic{theorem}}
\renewcommand\thelemma{A\theappen.\arabic{lemma}}
\renewcommand\theproposition{A\theappen.\arabic{proposition}}

\appen{$H^2_{\mathbf E}$ for $n=1$ antibracket}\label{appproof}

\subappen{General solution}

Let $\epsilon = \varepsilon +1$.
The conomology equation for antibracket can be represented in the form%
\begin{eqnarray}
d_{2}^{\mathrm{ad}}M_{2}(f,g,h)=-(-1)^{\epsilon (f)\epsilon
(h)}J_{M_{2},m_{0}}(z|f,g,h)=  \notag \\
\!\!\! -(-1)^{\epsilon (f) \epsilon (h)}\left( (-1)^{\epsilon
(f)\epsilon (h)}[[M_{2}(z|f,g),h(z)]+M_{2}(z|[f,g],h)]+\mathrm{%
cycle}(f,g,h)\right) =
0.  \label{5.2.1}
\end{eqnarray}

Introduce notation:%
\begin{equation*}
f(z)=f_{0}(x)+\xi f_{1}(x)=\check{f}_{0}(z)+\check{f}_{1}(z),\;\check{f}%
_{0}(z)=f_{0}(x),\;\check{f}_{1}(z)=\xi f_{1}(x).
\end{equation*}

Represent the forms $M_{1}(z|f)$ and $M_{2}(z|f,g)$ in the form
\begin{equation*}
M_{1}(z|f)=T_{(1)}(x|f_{0})+T_{(2)}(x|f_{1})+\xi \lbrack
T_{(3)}(x|f_{0})+T_{(4)}(x|f_{1})],
\end{equation*}%
\begin{eqnarray*}
&&M_{2}(z|f,g)=M_{(1)}(x|f_{0},g_{0})+M_{(2)}(x|f_{0},g_{1})-M_{(2)}(x|g_{0},f_{1})+M_{(3)}(x|f_{1},g_{1})+
\\
&&+\xi \lbrack
M_{(4)}(x|f_{0},g_{0})+M_{(5)}(x|f_{0},g_{1})-M_{(5)}(x|g_{0},f_{1})+M_{(6)}(x|f_{1},g_{1})],
\\
&&M_{(1,4)}(x|\varphi ,\phi )=M_{(1,4)}(x|\phi ,\varphi
),\;M_{(3,6)}(x|\varphi ,\phi )=-M_{(3,6)}(x|\phi ,\varphi ).
\end{eqnarray*}%
We have for $M_{2d}(z|f,g)=d_{1}^{\mathrm{ad}}M_{1}(f,g)$:
\begin{eqnarray}
&&M_{d|(1)}(x|\varphi ,\phi )=-T_{(3)}(x|\varphi )\partial _{x}\phi
(x)-T_{(3)}(x|\phi )\partial _{x}\varphi (x),\;M_{d|(4)}(x|\varphi ,\phi )=0,
\label{5.2.1a} \\
&&M_{d|(2)}(x|\varphi ,\phi )=\partial _{x}T_{(1)}(x|\varphi )\phi
(x)+T_{(4)}(x|\phi )\partial _{x}\varphi (x)-T_{(1)}(x|[\varphi ,\phi ]_{0}),
\label{5.2.1b} \\
&&M_{d|(3)}(x|\varphi ,\phi )=\partial _{x}T_{(2)}(x|\varphi )\phi
(x)-\partial _{x}T_{(2)}(x|\phi )\varphi (x)-T_{(2)}(x|[\varphi ,\phi ]_{1}),
\label{5.2.1c} \\
&&M_{d|(5)}(x|\varphi ,\phi )=\partial _{x}T_{(3)}(x|\varphi )\phi
(x)-T_{(3)}(x|\varphi )\partial _{x}\phi (x)-T_{(3)}(x|[\varphi ,\phi ]_{0}),
\label{5.2.1d} \\
&&M_{d|(6)}(x|\varphi ,\phi )=\partial _{x}T_{(4)}(x|\varphi )\phi
(x)-\partial _{x}T_{(4)}(x|\phi )\varphi (x)+T_{(4)}(x|\phi )\partial
_{x}\varphi (x)-  \notag \\
&&-T_{(4)}(x|\varphi )\partial _{x}\phi (x)-T_{(4)}(x|[\varphi ,\phi ]_{1}),
\label{5.2.1e} \\
&&\,[\varphi (x),\phi (x)]_{0}=\{\partial _{x}\varphi (x)\}\phi
(x),\;[\varphi (x),\phi (x)]_{1}=\{\partial _{x}\varphi (x)\}\phi
(x)-\varphi (x)\partial _{x}\phi (x).  \notag
\end{eqnarray}

It follows from $J_{M_{2},m_{0}}(z|\check{f}_{0},\check{g}_{0},\check{h}%
_{0})=0$ and $J_{M_{2},m_{0}}(z|\check{f}_{0},\check{g}_{0},\check{h}_{1})=0$ that %
\begin{eqnarray}
&&M_{(4)}(x|\varphi ,\phi )\partial _{x}\omega (x)+\mathrm{cycle}(\varphi
,\phi ,\omega )=0,  \label{5.2.2a} \\
&&M_{(4)}(x|\varphi ,\phi )\partial _{x}\omega (x)-\{\partial
_{x}M_{(4)}(x|\varphi ,\phi )\}\omega (x)+M_{(4)}(x|[\varphi ,\omega
]_{0},\phi )+M_{(4)}(x|\varphi ,[\phi ,\omega ]_{0})=0,  \notag
\end{eqnarray}%
\begin{eqnarray}
&&M_{(1)}(x|[\varphi ,\omega ]_{0},\phi )+M_{(1)}(x|\varphi ,[\phi ,\omega
]_{0})-\{\partial _{x}M_{(1)}(x|\varphi ,\phi )\}\omega (x)-  \notag \\
&&\,-M_{(5)}(x|\varphi ,\omega )\partial _{x}\phi (x)-M_{(5)}(x|\phi ,\omega
)\partial _{x}\varphi (x)=0.  \label{5.2.2c}
\end{eqnarray}

It follows from $d_{2}^{\mathrm{ad}}M_{2}(z|\check{f}_{0},\check{g}_{1},%
\check{h}_{1})=0$ that%
\begin{eqnarray}
&&M_{(5)}(x|\varphi ,\phi )\partial _{x}\omega (x)-\{\partial
_{x}M_{(5)}(x|\varphi ,\phi )\}\omega (x)+\{\partial _{x}M_{(5)}(x|\varphi
,\omega )\}\phi (x)-  \notag \\
&&\,-M_{(5)}(x|\varphi ,\omega )\partial _{x}\phi (x)-M_{(5)}(x|[\varphi
,\phi ]_{0},\omega )+M_{(5)}(x|[\varphi ,\omega ]_{0},\phi
)+M_{(5)}(x|\varphi ,[\phi ,\omega ]_{1})=0,  \label{5.2.2b}
\end{eqnarray}%
\begin{eqnarray}
&&\{\partial _{x}M_{(2)}(x|\varphi ,\omega )\}\phi (x)-\{\partial
_{x}M_{(2)}(x|\varphi ,\phi )\}\omega (x)-M_{(2)}(x|[\varphi ,\phi
]_{0},\omega )+  \notag \\
&&\,+M_{(2)}(x|[\varphi ,\omega ]_{0},\phi )+M_{(2)}(x|\varphi ,[\phi
,\omega ]_{1})+M_{(6)}(x|\phi ,\omega )\partial _{x}\varphi (x)=0.
\label{5.2.2e}
\end{eqnarray}

It follows from $d_{2}^{\mathrm{ad}}M_{2}(z|\check{f}_{1},\check{g}_{1},%
\check{h}_{1})=0$ that%
\begin{eqnarray}
&&M_{(6)}(x|\varphi ,\phi )\partial _{x}\omega (x)-\{\partial
_{x}M_{(6)}(x|\varphi ,\phi )\}\omega (x)-M_{(6)}(x|[\varphi ,\phi
]_{1},\omega )+  \notag \\
&&\,+\mathrm{cycle}(\varphi ,\phi ,\omega )=0,  \label{5.2.2d}
\end{eqnarray}%
\begin{equation}
-[\{\partial _{x}M_{(3)}(x|\varphi ,\phi )\}\omega (x)+M_{(3)}(x|[\varphi
,\phi ]_{1},\omega )+\mathrm{cycle}(\varphi ,\phi ,\omega )]=0.
\label{5.2.2f}
\end{equation}

I. Consider Eq. (\ref{5.2.2a}). As it was shown in \cite{JMP}, we find%
\begin{equation*}
M_{(4)}(x|\varphi ,\phi )=0.
\end{equation*}

II. Consider Eq. (\ref{5.2.2b}). As it was shown in \cite{JMP}, we find

\begin{equation*}
M_{(5)}(x|\varphi ,\phi )=M_{d|(5)}(x|\varphi ,\phi )+\mathrm{loc},
\end{equation*}%
where the expression for $M_{d|(5)}(x|\varphi ,\phi )$ is given by Eq. (\ref%
{5.2.1d}).

III. Consider Eq. (\ref{5.2.2c})..As it was shown in \cite{JMP}, we find%
\begin{equation*}
M_{(1)}(x|\varphi ,\phi )=M_{d|(1)}(x|\varphi ,\phi )+\mathrm{loc},
\end{equation*}%
where (the expression for $M_{d|(1)}(x|\varphi ,\phi )$ is given by Eq. (\ref%
{5.2.1a}).

IV. Consider Eq. (\ref{5.2.2d}). As it was shown in \cite{JMP}, we find

\begin{equation*}
M_{(6)}(x|\varphi ,\phi )=M_{d|(6)}(x|\varphi ,\phi )+\mathrm{loc},
\end{equation*}%
where the expression for $M_{d|(6)}(x|\varphi ,\phi )$ is given by Eq. (\ref%
{5.2.1e}).

V. Consider Eq. (\ref{5.2.2e}). As it was shown in \cite{JMP}, we find

\begin{eqnarray*}
M_{(2)}(x|\varphi ,\phi ) &=&M_{(2)8}(x|\varphi ,\phi )+M_{d|(2)}(x|\varphi
,\phi )+\mathrm{loc}, \\
M_{(2)8}(x|\varphi ,\phi ) &=&\sum_{q=0,q\neq 1}^{Q}M_{7}^{q}(x|\{\partial
^{q}\varphi \}\phi ),\;\partial _{x}\hat{M}_{7}^{q}(x|\varphi )=0,
\end{eqnarray*}%
where the expression for $M_{d|(2)}(x|\varphi ,\phi )$ is given by Eq. (\ref%
{5.2.1b}).

For $M_{(2)8}(x|\varphi ,\phi )$ we obtain an equation
\begin{eqnarray*}
&&\{\partial _{x}M_{(2)8}(x|\varphi ,\omega )\}\phi (x)-\{\partial
_{x}M_{(2)8}(x|\varphi ,\phi )\}\omega (x)-M_{(2)8}(x|[\varphi ,\phi
]_{0},\omega )+ \\
&&\,+M_{(2)8}(x|[\varphi ,\omega ]_{0},\phi )+M_{(2)8}(x|\varphi ,[\phi
,\omega ]_{1})=\mathrm{loc}.
\end{eqnarray*}

Let
\begin{equation*}
x\cap \lbrack \mathrm{supp}(\varphi )\cup \mathrm{supp}(\phi )\cup \mathrm{%
supp}(\omega )]=\varnothing .
\end{equation*}%
We obtain
\begin{equation*}
\hat{M}_{(2)8}(x|[\varphi ,\phi ]_{0},\omega )-\hat{M}_{(2)8}(x|[\varphi
,\omega ]_{0},\phi )-\hat{M}_{(2)8}(x|\varphi ,[\phi ,\omega ]_{1})=0
\end{equation*}%
or
\begin{equation}
\sum_{q=0,q\neq 1}^{Q}\hat{M}_{7}^{q}(x|\{\partial ^{q}(\partial \varphi
\phi )\}\omega -\{\partial ^{q}(\partial \varphi \omega )\}\phi -\{\partial
^{q}\varphi \}[\partial \phi \omega -\phi \partial \omega ])=0.
\label{5.2.3}
\end{equation}%
Let $\varphi (x)\rightarrow e^{px}\varphi (x)$\thinspace and $\phi
(x)\rightarrow e^{kx}$, $\omega (x)\rightarrow e^{-(p+k)x}$ for $x\in $ $%
\mathrm{supp}\varphi $.

Consider the terms of highest order in $p$, $k$ in Eq. (\ref{5.2.3}),
\begin{equation*}
\lbrack p(p+k)^{Q}-p(-k)^{Q}-(p+2k)p^{Q}]\hat{M}_{7}^{Q}(x|\varphi
)=0\;\Longrightarrow
\end{equation*}%
\begin{eqnarray*}
&&\hat{M}_{7}^{q}(x|\varphi )=0,\;q\neq 0,2\;\Longrightarrow
\;M_{(2)8}(x|\varphi ,\phi )=M_{7}^{2}(x|\{\partial ^{2}\varphi \}\phi
)+M_{7}^{0}(x|\varphi \phi )+\mathrm{loc}, \\
&&\,\partial \hat{M}_{7}^{2}(x|\varphi )=\partial \hat{M}_{7}^{0}(x|\varphi
)=0.
\end{eqnarray*}

Consider the terms of the second order in $p$, $k$ in eq. (\ref{5.2.3}) (the
terms of third order are identically cancelled),%
\begin{eqnarray*}
&&(p^{2}+2pk)\hat{M}_{7}^{2}(x|\partial \varphi )=0\;\Longrightarrow
\;\partial _{x}\hat{m}_{7}^{2}(x|y)=\hat{m}_{7}^{2}(x|y)\overleftarrow{%
\partial _{y}}=0\;\Longrightarrow \\
&&m_{7}^{2}(x|y)=c_{5}+2c_{6}\theta (x-y)+\mathrm{loc},
\end{eqnarray*}%
where%
\begin{eqnarray*}
&&M_{7}^{2}(x|\varphi )=\int dym_{7}^{2}(x|y)\varphi (y)=c_{5}\int dy\varphi
(y)+c_{6}\int dy\theta (x-y)\varphi (y)+\mathrm{loc}\;\Longrightarrow \\
&&\,M_{(2)8}(x|\varphi ,\phi )=c_{5}\int dy\{\partial ^{2}\varphi (y)\}\phi
(y)+ \\
&&+2c_{6}\int dy\theta (x-y)\{\partial ^{2}\varphi (y)\}\phi
(y)+M_{7}^{0}(x|\varphi \phi )+\mathrm{loc}.
\end{eqnarray*}

It follows from eq. (\ref{5.2.3})%
\begin{equation*}
\hat{M}_{7}^{0}(x|\varphi \partial \phi \omega -\varphi \phi \partial \omega
)=0\;\Longrightarrow \;M_{7}^{0}(x|\varphi )=\mathrm{loc}.
\end{equation*}

Finally, we have
\begin{eqnarray*}
&&M_{(2)}(x|\varphi ,\phi )=c_{5}\tilde {\mu} _{2|5}(x|\varphi ,\phi )+2c_{6}\tilde{\mu}
_{2|6}(x|\varphi ,\phi )+ M_{d|(2)}(x|\varphi ,\phi )+\mathrm{loc}, \\
&&\tilde{\mu} _{2|5}(x|\varphi ,\phi )=\int dy\{\partial ^{2}\varphi (y)\}\phi
(y),\;\tilde{\mu} _{2|6}(x|\varphi ,\phi )=\int dy\theta (x-y)\{\partial ^{2}\varphi
(y)\}\phi (y),
\end{eqnarray*}
or, after equivalent transformations and notation changing
\begin{eqnarray*}
&&M_{(2)}(x|\varphi ,\phi )=c_{5}\mu _{2|5}(x|\varphi ,\phi )+ c_{6}\mu
_{2|6}(x|\varphi ,\phi )+ M_{d|(2)}(x|\varphi ,\phi )+\mathrm{loc}, \\
&&\mu _{2|5}(x|\varphi ,\phi )=\int dy\{\partial \varphi (y)\}\partial \phi
(y),\;\mu _{2|6}(x|\varphi ,\phi )=\int dy\theta (x-y)\{\partial \varphi
(y)\}\partial_y\phi (y),
\end{eqnarray*}

VI. Consider Eq. (\ref{5.2.2f}). As it was shown in \cite{JMP}, we find%
\begin{eqnarray*}
M_{(3)}(x|\varphi ,\phi ) &=&c_{1}\mu _{2|1}(x|\varphi ,\phi )+c_{2}\mu
_{2|2}(x|\varphi ,\phi )+M_{d|(3)}(x|\varphi ,\phi )+\mathrm{loc}, \\
\mu _{2|1}(x|\varphi ,\phi ) &=&\int dy[\partial _{y}^{3}\varphi (y)]\phi
(y), \\
\mu _{2|2}(x|\varphi ,\phi ) &=&\int dy\theta (x-y)[\{\partial
_{y}^{3}\varphi (y)\}\phi (y)-\varphi (y)\partial _{y}^{3}\phi (y)].
\end{eqnarray*}

Introduce two forms $m_{2|a}(z|f,g)$, $a=1,2$, $\epsilon _{m_{2|a}}=1$,%
\begin{eqnarray*}
&&m_{2|1}(z|f,g)=\int du(-1)^{\epsilon (f)}[\partial
_{y}^{3}f(u)]\partial _{\eta }g(u), \\
&&m_{2|2}(z|f,g)=\int du\theta (x-y)(-1)^{\epsilon (f)}\left\{ [\partial
_{y}^{3}f(u)]\partial _{\eta }g(u)+(-1)^{\varepsilon (g)}[\partial _{\eta
}f(u)]\partial _{y}^{3}g(u)\right\} - \\
&&\,-x\left\{ [\partial _{x}^{2}\partial _{\xi }f(z)]\partial _{x}\partial
_{\xi }g(z)-[\partial _{x}\partial _{\xi }f(z)]\partial _{x}^{2}\partial
_{\xi }g(z)\right\} .
\end{eqnarray*}

These forms have the properties:%
\begin{eqnarray*}
&&m_{2|a}(z|\check{f}_{0},\check{g}_{0})=m_{2|a}(z|\check{f}_{1},\check{g}%
_{0})=m_{2|a}(z|\check{f}_{0},\check{g}_{1})=0, \\
&&m_{2|1}(z|\check{f}_{1},\check{g}_{1})=\int dy[\partial
^{3}f_{1}(y)]g_{1}(y)=\mu _{2|1}(x|f_{1},g_{1}), \\
&&m_{2|2}(z|\check{f}_{1},\check{g}_{1})=\int dy\theta (x-y)[\{\partial
_{y}^{3}f_{1}(y)\}g_{1}(y)-f_{1}(y)\partial _{y}^{3}g_{1}(y)]- \\
&&-x[\{\partial _{x}^{2}f_{1}(x)\}\partial _{x}g_{1}(x)-\{\partial
_{x}f_{1}(x)\}\partial _{x}^{2}g_{1}(x)]=\mu _{2|2}(x|f_{1},g_{1})+\mathrm{%
loc}.
\end{eqnarray*}%
\begin{eqnarray*}
&&m_{2|a}(z|g,f)=-(-1)^{\epsilon (f)\epsilon
(g)}m_{2|a}(z|f,g), \\
&&d_{2}^{\mathrm{ad}}m_{2|a}(z|f,g,h)=0.
\end{eqnarray*}

Introduce two forms $m_{2|a}(z|f,g)$, $a=5,6$, $\epsilon _{m_{2|a}}=0$,%

\begin{eqnarray*}
&&m_{2|5}(z|f,g)=\int du (-1)^{\epsilon(f)}\partial_y f(u)\partial_y g(u), \\
&&m_{2|6}(z|f,g)=\int du\theta(x-y)(-1)^{\epsilon(f)}\partial_y f(u)\partial_y g(u).
\end{eqnarray*}

These forms have the properties:%

\begin{eqnarray*}
&&m_{2|a}(z|\check{f}_0,\check{g}_0)=m_{2|a}(z|\check{f}_1,\check{g}_1)=0, \\
&&m_{2|5}(z|\check{f}_0,\check{g}_1)=
\int dy[\partial_y f_{0}(y)]\partial_y g_1(y)=\mu _{2|5}(x|f_0,g_1), \\
&&m_{2|6}(z|\check{f}_{0},\check{g}_{1})=
\int dy\theta (x-y)[\partial_y f_0(y)]\partial_y g_1(y)
=\mu_{2|6}(x|f_{0},g_1).
\end{eqnarray*}%
\begin{eqnarray*}
&&m_{2|a}(z|g,f)=-(-1)^{\epsilon (f)\epsilon
(g)}m_{2|a}(z|f,g), \\
&&d_{2}^{\mathrm{ad}}m_{2|a}(z|f,g,h)=0.
\end{eqnarray*}

So, we obtained
\begin{eqnarray*}
&&M_{2}(z|f,g)=c_{1}m_{2|1}(x|f,g)+c_{2}m_{2|2}(x|f,g)+c_{5}m_{2|5}(x|f,g)+
\\
&&+c_{6}m_{2|6}(x|f,g)+d_{1}^{\mathrm{ad}}M_{1|1}(z|f,g)+M_{2\mathrm{loc}%
}(z|f,g).
\end{eqnarray*}%
The local form $M_{2\mathrm{loc}}(z|f,g)$ satisfies the equation $d_{2}^{%
\mathrm{ad}}M_{2\mathrm{loc}}(z|f,g,h)=0$, the solution of which, as it was
shown in \cite{JMP}, is%
\begin{eqnarray*}
&&M_{2\mathrm{loc}}(z|f,g)=c_{3}m_{2|3}(x|f,g)+c_{4}m_{2|4}(x|f,g)+d_{1}^{%
\mathrm{ad}}M_{1|2}(z|f,g), \\
&&m_{2|3}(x|f,g)=(-1)^{\varepsilon (f)}\{(1-N_{\xi })f(z)\}(1-N_{\xi
})g(z),\;\varepsilon _{m_{2|3}}=0, \\
&&m_{2|4}(x|f,g)=(-1)^{\varepsilon (f)}[\Delta f(z)]\hat{l}_{z}g(z)+[\hat{l}%
_{z}f(z)]\Delta g(z),\;\varepsilon _{m_{2|4}}=1.
\end{eqnarray*}

Finally, we find: general solution of eq. (\ref{5.2.1}) is%
\begin{equation*}
M_{2}(z|f,g)=\sum_1^6 c_{i}m_{2|i}(x|f,g)+d_{1}^{\mathrm{ad}}M_{1}(z|f,g).%
\end{equation*}

\subappen{Adjoint Cohomology}

Let $D$ denotes $\mathbf D_1$.
We will say that the form $M(f,g,...)$ is compact
and we will write $M=\mathrm{comp}$ if
$M(f,g,...)\in D$ for any $f,g,... \in D$.

\bigskip Here we prove an useful

{\bf Statement} The form
$
M_{2}(z|f,g)
$ %
is compact iff $c_{1}=c_{2}=c_{5}=c_{6}=0$.and $d_{1}^{%
\mathrm{ad}}M_{1}(z|f,g)=\mathrm{comp}.$

Proof.

We must solve the equations%
\begin{eqnarray}
c_{1}m_{2|1}(x|f,g)+c_{2}m_{2|2}(x|f,g)+M_{1}(z|\{f,g\}) &=&\mathrm{comp}%
,\;\varepsilon _{M_{1}}=1,  \label{5.2.4} \\
c_{5}m_{2|5}(x|f,g)+c_{6}m_{2|6}(x|f,g)+M_{1}(z|\{f,g\}) &=&\mathrm{comp}%
,\;\varepsilon _{M_{1}}=0.  \label{5.2.5}
\end{eqnarray}

First, consider eq. (\ref{5.2.4}).

It must be $\varepsilon _{M_{1}}=1$, such that we have%
\begin{eqnarray*}
M_{1}(z|f) &=&\int dum_{1}(z|u)f(u),\;m_{1}(z|u)=\mu (x|y)+\xi \eta \nu
(x|y)\;\Longrightarrow  \\
M_{1}(z|f) &=&\int dy^{0}\mu (x|y)f_{1}(y)-\xi \int dy\nu (x|y)f_{0}(y),
\end{eqnarray*}%
and%
\begin{eqnarray}
&&c_{1}m_{2|1}(z|f,g)+c_{2}m_{2|2}(z|f,g)+\int dy\mu (x|y)[f_{1}^{\prime
}(y)g_{1}(y)-f_{1}(y)g_{1}^{\prime }(y)]=  \notag \\
&&\,=\mathrm{comp},  \label{n1.1}
\end{eqnarray}%
\begin{equation}
\int dy\nu (x|y)[f_{0}^{\prime }(y)g_{1}(y)-f_{1}(y)g_{0}^{\prime }(y)]=%
\mathrm{comp}.  \label{n1.2}
\end{equation}

Consider eq. (\ref{n1.2}). Choosing $g_{0}(y)=y$ for $y\in \mathrm{supp}f_{1}
$ and $f_{0}(y)=1$ for $y\in \mathrm{supp}g_{1}$, we obtain

\begin{equation*}
\int dy\nu (x|y)f(y)=\mathrm{comp}\text{, }\forall f\in D.
\end{equation*}

Turn to eq. (\ref{n1.1}).

i) Choosing $f_{1}(y)=1$ for $y\in \mathrm{supp}g_{1}$, we obtain%
\begin{eqnarray*}
&&\int dy\mu (x|y)g^{\prime }(y)=\mathrm{comp},\;\forall g\in
D\;\Longrightarrow  \\
&&\,\Longrightarrow c_{1}m_{2|1}(z|f,g)+c_{2}m_{2|2}(z|f,g)-2\int dy\mu
(x|y)f_{1}(y)g_{1}^{\prime }(y)=\mathrm{comp}.
\end{eqnarray*}%
Further choosing $g_{0}(y)=y$ for $y\in \mathrm{supp}f_{1}$, we find finally%
\begin{equation*}
\int du^{0}\mu (x|u^{0})f(u^{0})=\mathrm{comp},\;\forall f\in D,
\end{equation*}%
and as a consequence%
\begin{equation*}
c_{1}m_{2|1}(z|f,g)+c_{2}m_{2|2}(z|f,g)=\mathrm{comp}.
\end{equation*}%
Let $x\rightarrow -\infty .$ We obtain $c_{1}m_{2|1}(z|f,g)=\mathrm{comp}$ $%
\Longrightarrow $ $c_{1}=0$ $\Longrightarrow $ $c_{2}=0$.

Now, consider eq. (\ref{5.2.5}).

Since $\varepsilon _{M_{1}}=0$, we have%
\begin{equation*}
M_{1}(z|f)=\int dy^{0}\mu (x|y)f_{0}(y)-\xi \int dy\nu (x|y)f_{1}(y),
\end{equation*}%
and%
\begin{equation}
c_{5}\mu _{2|5}(x|f_{0},g_{1})+c_{6}\mu _{2|6}(x|f_{0},g_{1})+\int dy\mu
(x|y)f_{0}^{\prime }(y)g_{1}(y)=\mathrm{comp},  \label{n1.3}
\end{equation}%
\begin{equation}
\int dy\nu (x|y)[f_{1}^{\prime }(y)g_{1}(y)-f_{1}(y)g_{1}^{\prime }(y)]=%
\mathrm{comp}.  \label{n1.4}
\end{equation}

Setting $g_{1}(y)=1$ for $y\in \mathrm{supp}f_{1}$ in eq. (\ref{n1.4}), we
find $\int dy\nu (x|y)f_{1}^{\prime }(y)=\mathrm{comp}$, $\forall
f_{1}(y)\in D$ $\Longrightarrow $ $\int dy\nu (x|y)f_{1}(y)g_{1}^{\prime
}(y)]=\mathrm{comp}.$ Choosing $g_{1}(y)=y$ for $y\in \mathrm{supp}f_{1}$%
, we obtain%
\begin{equation*}
\int dy\nu (x|y)f(y)=\mathrm{comp},\;\forall f(y)\in D.
\end{equation*}%
Now, setting $f_{0}(y)=y$ for $y\in \mathrm{supp}g_{1}$ in eq. (\ref{n1.3}),
we find%
\begin{eqnarray*}
&&\int dy\mu (x|y)g(y)=\mathrm{comp},\;\forall g(y)\in D\;\Longrightarrow  \\
&&\,\Longrightarrow c_{5}\mu _{2|5}(x|f_{0},g_{1})+c_{6}\mu
_{2|6}(x|f_{0},g_{1})=\mathrm{comp}\;\Longrightarrow \;c_{5}=c_{6}=0.
\end{eqnarray*}

As a consequence, all forms $m_{2|i}(z|f,g)$, $i=1,2,...,6$, are independent
nontrivial cohomology.


\begin{thebibliography}{99}


\bibitem{BV1}
{\it Batalin I.A., Vilkovisky G.A.,} Phys. Lett., {\bf 120B}, 166 (1983).

\bibitem{BV2}
{\it I.A. Batalin and G.A. Vilkovisky,} J. Math. Phys., {\bf 26}, 172 (1985).

\bibitem{reports}
{\it Gomis J., Paris J., Samuel S.,} Antibrackets, antifields and gauge theory
quantization, Phys. Rep., {\bf 259} 1--145 (1995).

\bibitem{books}{\it D.M.Gitman and I.V.Tyutin,}
Quantization of Fields with Constraints,
(Springer--Verlag, 1990).

\bibitem{HenTei}
{\it Henneaux M. and Teitelboim C.,} Quantization of Gauge Systems,
Princeton University Press, Princeton, 1992.

\bibitem{Leites} {\it D.~A.~Leites and I.~M.~Shchepochkina,}
How to quantize the antibracket, Theor.~Math.~Phys., {\bf
126}, 281--306 (2001).

\bibitem{Schei97} {\it M.~Scheunert and R.~B.~Zhang,} J.Math.Phys., {\bf 39},
5024--5061 (1998); q-alg/9701037.

\bibitem{SKT1} {\it S.~E.~Konstein, A.~G.~Smirnov and I.~V.~Tyutin,}
Cohomologies of the Poisson superalgebra,
Teor.~Mat.~Fiz., {\bf 143},625 (2005);
hep-th/0312109.

\bibitem{JMP} {\it S.~E.~Konstein, and I.~V.~Tyutin,}
Deformations and central extensions of the antibracket Superalgebra,
Journal of Mathematical Physics, 49, 072103 (2008).

\bibitem{poi} {\it S.~E.~Konstein, and I.~V.~Tyutin,}
The deformations of nondegenerate constant Poisson bracket
with even and odd deformation parameters, arXiv: 1001.1776 [hep-th]


\end{thebibliography}
\end{document}